\documentclass[
reprint,
amsmath,
amssymb,
aps,
]{revtex4-2}

\usepackage{graphicx}
\usepackage{dcolumn}
\usepackage{bm}
\usepackage{hyperref}
\usepackage{upgreek}
\usepackage{eqnarray,amsmath}
\usepackage{float}
\usepackage{xcolor}
\usepackage{svg}
\usepackage{soul}

\renewcommand{\st}[1]{}

\begin{document}


\title{Dispersive and dissipative coupling of photon Bose-Einstein condensates}%

\author{Chris Toebes}
\author{Mario Vretenar}
\author{Jan Klaers}
\email{j.klaers@utwente.nl}

\affiliation{
 Adaptive Quantum Optics (AQO), MESA$^+$ Institute for Nanotechnology, University of Twente, PO Box 217, 7500 AE Enschede, Netherlands
}


\date{\today}
             

\maketitle

\onecolumngrid
\st{\textbf{Synchronizing coherent states of light has been a long standing topic of research. In recent years, a new paradigm has arrived where this synchronization process can be used to solve computationally hard problems. Many platforms have been suggested for implementing such computations. In this work, we analyze and characterize the synchronization process of two photon Bose-Einstein condensates (pBECs) using two distinct methods, disspative and dispersive coupling, and evaluate the pBEC system's potential for all-optical computation. We find a similar coupling strength of 10-20 GHz for both measurements. The system is also analyzed under imbalanced pumping conditions and significant differences in the behaviour of both coupling types are uncovered with both time-resolved and time-averaged measurements.}}


{\color{black}The synchronization of coherent states of light has long been an important subject of basic research and technology. Recently, a new concept for analog computers has emerged where this synchronization process can be exploited to solve computationally hard problems – potentially faster and more energy-efficient than what can be achieved with conventional computer technology today. The unit cell of such systems consists of two coherent centers that are coupled to one another in a controlled manner.
Here, we experimentally characterize and analyze the synchronization process of two photon Bose-Einstein condensates, which are coupled to one another, either dispersively or dissipatively. We show that both types of coupling are robust against a detuning of the condensate frequencies and show similar time constants in establishing mutual coherence. Significant differences between these couplings arise in the behaviour of the condensate populations under imbalanced optical pumping. The combination of these two types of coupling extends the class of physical models that can be investigated using analog simulations.}
\vspace{5mm}
$\,$

\twocolumngrid

\section*{Introduction}
\st{Analog simulations have a long tradition in physics as a computational \st{aid} {\color{black} tool [Refs?]}.}  {\color{black}Analog simulations are of great importance in physics whenever a physical problem cannot be solved exactly or numerically due to its mathematical complexity.} In recent years, \st{analog} {\color{black}such} simulations have gained attention as a possible method to outperform the current standard of digital silicon computers {\color{black} in specific computational tasks}. Of particular interest \st{to this work} {\color{black}in this regard} is optical spin-glass simulation \cite{nixon2013observing,marandi2014network,mcmahon2016fully,inagaki2016large,inagaki2016coherent,ohadi2016nontrivial,berloff2017realizing,lagoudakis2017polariton,kalinin2018global,ohadi2018synchronization,kalinin2019polaritonic,pal2020rapid,parto2020nanolaser,parto2020realizing,alyatkin2020optical,kalinin2020polaritonic,kalinin2020toward,vretenar2021controllable}. Here, \st{the goal is to build} a system of coupled coherent states of light {\color{black} is used to simulate classical spin models such as the Ising or XY model. The basic idea behind these simulators is to map the Hamiltonian of a classical spin model to the gain function of an optical system. A mode competition is then set in motion by optical pumping, which leads to a sampling of low-energy configurations of the simulated spin model.} \st{that emulates an array of coupled (magnetic) spins and finds the energetic ground state, which is an NP-hard optimization problem.}  This can provide{\color{black}, for example,} interesting insight into physics of magnetic systems, such as the BKT-transition \cite{berezinskii1972destruction,kosterlitz1973ordering}\st{, but solving this ground-state energy problem is also} {\color{black}. However, the potential applications of such simulators go far beyond physics, since finding the lowest energy configuration of a spin model with sufficiently general couplings (spin-glass problem) is known to be} a NP-hard optimization problem \cite{barahona1982computational,cubitt2016complexity}{\color{black}.} A device that can produce solutions to the spin-glass problem quickly and {\color{black}energy-}efficiently {\color{black}will also be highly useful for providing} \st{can therefore also provide} solutions to {\color{black} mathematically equivalent} optimization tasks that are ubiquitous in modern society, like job scheduling and the knapsack problem \cite{lucas2014ising}. {\color{black} For the Ising model, a first generation of analog spin-glass simulators has been realized using networks of superconducting qubits \cite{johnson2011quantum,boixo2014evidence} and optical parametrical oscillators \cite{marandi2014network,mcmahon2016fully,inagaki2016large,inagaki2016coherent}. For the XY model, the proposed physical platforms are based on superconducting qubits \cite{king2018observation}, lasers \cite{nixon2013observing,pal2020rapid,parto2020nanolaser,parto2020realizing,honari2020mapping,honari2020optical}, atomic Bose-Einstein condensates (BEC) \cite{struck2013engineering}, and optical Bose-Einstein condensates \cite{ohadi2016nontrivial,berloff2017realizing,lagoudakis2017polariton,kalinin2018simulating,ohadi2018synchronization,kalinin2019polaritonic,alyatkin2020optical,kalinin2020polaritonic}.

An optical XY spin-glass simulator consists of a lattice of coherent states of light, which represent the XY spins, i.e. angular degrees of freedom $\theta_{i}\in[0,2\pi)$ in the simulated model. By physically coupling these states, one tries to map the Hamiltonian of the simulated XY model, i.e.  $H_{\mathrm{XY}}=-\Sigma J_{ij}\cos{(\theta_{i}-\theta_{j})}$,  to the gain function of the simulator system. It is important to distinguish the physical couplings in the simulator system from the coupling constants in the simulated model. While the couplings in the simulated XY model are real-valued ($J_{ij}\in \mathbb{R}$), the physical couplings between the coherent states in the simulator system are generally described by complex-valued parameters since the simulators typically operate under non-equilibrium conditions, i.e. they are determined by gain and loss.}
\st{The basis of our platform is the coupled photon Bose-Einstein condensate (BEC) system
in which we aim to implement the specific version of the spin-glass problem called the XY-model. The Hamiltonian that governs this system is $H_{\mathrm{XY}}=-\Sigma J_{ij}\cos{(\theta_{i}-\theta_{j})}$, where $J_{ij}$ is the coupling between spins (condensates) and $\theta_{i,j}$ is the angle of the spin (phase of the condensate). An essential difference between the standard XY-model and the photon BEC system is the fact that in the standard XY-model $J_{ij}$ is a real valued number, while in the photonic variant it can also be complex.} {\color{black} The c}\st{C}oupling in these {\color{black} simulator} systems is typically divided into two groups, namely dispersive \st{{\color{black}(real-valued)}} and dissipative \cite{ding2019dispersive,ding2019mode}. \st{\color{black}(purely imaginary)} \st{coupling.} Dispersive coupling is caused by direct particle exchange between the \st{condensates} {\color{black}coherent centers}. Here, the coupling constant is real-valued and in turn leads to splitting in the eigenfrequencies of the coupled system. In contrast, dissipative coupling is caused by the state of one \st{condensate} {\color{black}coherent center} affecting the loss rate of another \st{condensate} and therefore has an imaginary coupling constant. This leads to a splitting in the imaginary component of the eigenfrequencies of the coupled system. 

The nature \st{and strength} of the couplings between \st{individual} {\color{black}the} coherent \st{states} {\color{black}centers} dictate{\color{black}s} the overall dynamics of the {\color{black}simulator} system. {\color{black}The couplings not only define a specific gain function but also determine whether the system is aiming for fixed points in the configuration space.} Investigating the fundamental differences between dispersive and dissipative coupling is {\color{black}therefore} critical to understanding and improving optical spin-glass simulators {\color{black}and is the basis for extending their simulation capabilities to a larger class of physical models}. \st{In this paper, we study a two-condensate system with two distinct coupler designs. These designs exhibit either predominantly dispersive or predominantly dissipative coupling.} {\color{black}In this work, we investigate a system of two coupled photon Bose-Einstein condensates \cite{klaers2010bose,klaers2010thermalization,klaers2012statistical,kirton2013nonequilibrium,schmitt2015thermalization,dung2017variable,walker2018driven,greveling2018density,gladilin2020classical}, in which the flow of the photons is controlled in such a way that either a predominantly dispersive or a dissipative coupling of the condensates is created.} We find similar coupling strengths \st{in the order of 10-20\,GHz for both coupler designs} {\color{black} of several 10 GHz for both types of coupling}. A time-resolved measurement \st{also shows} {\color{black}reveals} the fast dynamics in these systems, {\color{black}e.g.,} \st{rapid} condensation and {\color{black}the establishment of} coherence \st{establishments in $< 1 \,$ns} {\color{black}within several 100 ps}. \st{We also investigate the influence of strong imbalanced illumination of the condensate site and find that it leads to a delay in condensation of the second condensate site. The dispersive coupler can also exhibit coherent oscillations between the symmetric and antisymmetric state in certain gain regimes.} {\color{black}A fundamental difference in the physical behavior of the two types of coupling arises in the case of unbalanced optical pumping.} 

\section*{Results}

\begin{figure*}[]
\begin{center}
  \includegraphics[width=18cm]{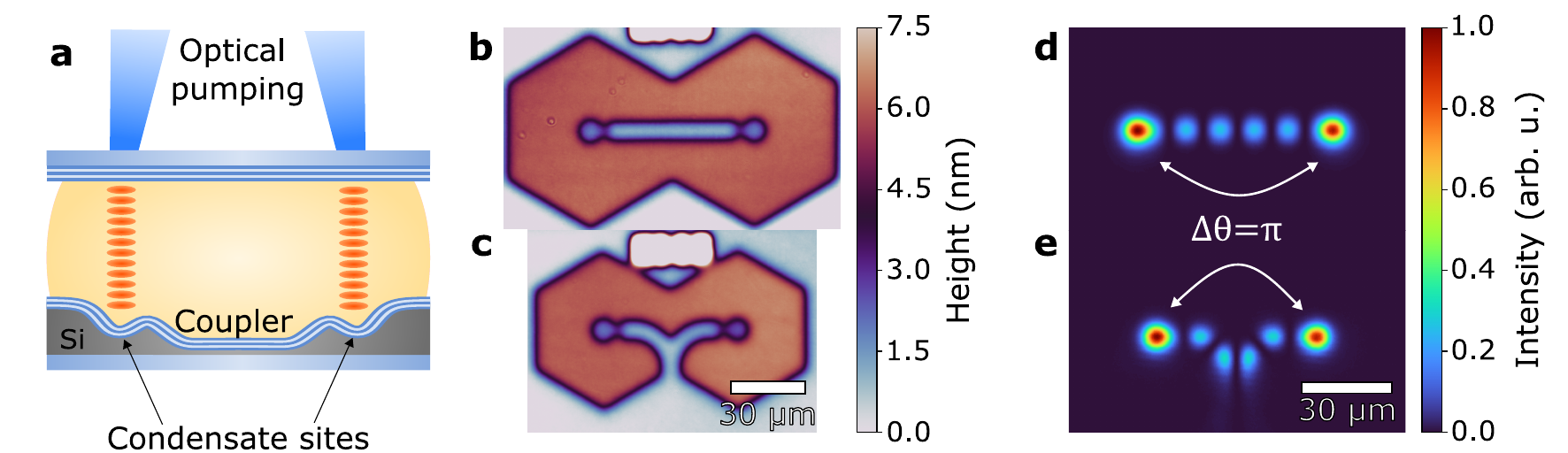}
  \caption{\textbf{$\mid$ Overview of the experiment and investigated coupling structures}. \textbf{a} Schematic representation (cross section) of the microcavity. The cavity consists of two highly reflective mirrors and is filled with a rhodamine 6G dye solution. A nanostructured mirror creates two circular confining potential, where local photon Bose-Einstein condensates form after local excitation with an optical pump beam. The coupling is determined by the potential landscape between the condensates.  \textbf{b,c} Height map of the structure for the direct and Y-coupling potential, respectively. \textbf{d,e} Measured photon densities under symmetric excitation for the direct coupler and the Y-coupler, respectively. These images are acquired by imaging the light that leaves the microcavity onto a camera.}
  \label{fig_cavity}
\end{center}
\end{figure*}

Our experimental setup consists of a high-finesse optical microcavity filled with rhodamine 6G dissolved in ethylene glycol {\color{black} as optically active medium}, see Fig. \ref{fig_cavity}a. One of the cavity mirrors is fixed in position, while the other is controlled by three piezo actuators that allow for precise adjustment of the cavity length and tilt.  The frequency-dependent absorption and emission coefficients of the dye follow the Kennard-Stepanov law: $B_{12}(\omega)/B_{21}(\omega)=\exp[\hbar(\omega\hspace{-0.5mm}-\hspace{-0.5mm}\omega_{\text{zpl}})/kT]$, where $\omega_{\text{zpl}}$ is the zero-phonon line of the dye. \st{ Here,} {\color{black}This means that} the absorption $B_{12}(\omega)$ and emission $B_{21}(\omega)$ are related by a Boltzmann factor. The high-reflectivity mirrors confine the photons and allow for repeated absorption and re-emission by the dye molecules before leaving the cavity. This brings the photon gas in thermal equilibrium with the dye and allows for the formation of a BEC at room temperature \cite{klaers2010thermalization,klaers2010bose}. When the separation between the microcavity mirrors is sufficiently small, the free spectral range can \st{be larger than} {\color{black}become comparable to} the emission bandwidth of the dye. \st{This results in only one possible longitudinal mode which is conserved during absorption-emission cycles. Therefore, the dynamics of the photon gas become two-dimensional.} {\color{black}This leads to a conservation of the longitudinal mode number of the photons when interacting with the optical medium, which effectively makes the photon gas two-dimensional.} The energy of \st{the} {\color{black}a} photon \st{gas} can then be approximated by 
\begin{equation}
\label{teq0}
E\simeq \frac{mc^2}{n_0^2}+\frac{(\hbar k_r)^2}{2m} - \frac{m c^2}{n_0^2}  \frac{\Delta d}{D_0} ,
\end{equation}
where $m$ denotes the effective photon mass, $n_0$ is refractive index of the dye medium,  $k_r$ is the transversal wavenumber, $D_0$ is the mean mirror separation and $\Delta d$ is \st{the} {\color{black}a} (small) change in mirror separation over the surface. The first term corresponds to the rest energy of the photon, the second term corresponds to the kinetic energy and the final term is equivalent to the potential energy. 
\st{However, we find that even for a larger mirror separation the system can still behave two-dimensionally as long as there is only one longitudinal mode with a significantly lower loss than the others.}
In this experiment we use two methods for controlling the potential energy term. The first is a \st{direct-laser-written microstructuring technique} {\color{black}nanostructuring technique based on direct-laser writing} that generates a well defined height profile on the mirror \cite{kurtscheid2020realizing}. The second is precise adjustment of the tilt angle between the two cavity mirrors {\color{black}by piezo-electric actuation}. The tilt translates into a height (hence potential) gradient in the cavity. 

{\color{black}{\bf Dispersive and dissipative coupling.}} \st{The coupling structures under investigation}{\color{black}The nanostructured mirrors that are used to create and investigate dispersive and dissipative couplings between two photon BECs} are shown in Fig. \ref{fig_cavity}b and \ref{fig_cavity}c. {\color{black}For the sake of simplicity, we refer to these structures as direct coupling and Y-coupling structure.} The direct coupler (Fig. \ref{fig_cavity}b) \st{is manufactured in a way that is similar to} {\color{black}has been introduced in} our previous work and can be considered as photonic Josephson junction \cite{vretenar2021controllable}. \st{It is designed to function as a dispersive coupler} {\color{black}It realizes (predominantly) dispersive coupling between condensates}. \st{In this structure,} {\color{black}Its height profile creates} two circular confining potentials \st{serve as condensate sites} {\color{black}that can host localized photon Bose-Einstein condensates}. Optically pumping these sites induces a \st{local} high {\color{black}local} chemical potential, leading to the formation of a photon BEC. The condensate\st{s} {\color{black}locations} are connected by a \st{waveguide with} {\color{black}waveguiding potential at} a lower potential {\color{black}energy}. \st{Light}{\color{black}Photons} that leak\st{s} from one of the condensates {\color{black}gain kinetic energy as they enter this waveguiding potential and} \st{can therefore} form a traveling wave towards the other condensate. The phase difference that the light accumulates while traveling through the waveguide determines whether the condensate{\color{black}s couple} \st{coupling is ferromagnetic (}in phase\st{)} or \st{antiferromagnetic (}in antiphase\st{)}. \st{Fig.} {\color{black}Figure} \ref{fig_cavity}d
shows an example of the \st{light emitted from the cavity for an antiferromagnetically coupled system for the direct and Y-coupler, respectively. The overall structure is enclosed in a high potential barrier to confine the light transversally.} {\color{black} photon density inside the microcavity plane obtained by observing the light that is transmitted through one of the cavity mirrors. The shown density pattern indicates an antisymmetric wavefunction that corresponds to an antiphase coupling of the two condensates.}
\st{For the Y-coupler in 
, the confinement at the condensate sites is designed to be similar to the direct coupler. The main difference is in the design of the waveguides. In the Y-coupler the condensates are not directly connected. Each condensate has a waveguide that arcs outward onto a flat region of the mirror surface. Light that travels from one condensate through the waveguide can leave onto the plane as a travelling wave. This creates an additional loss channel that depends on the phase relation between the condensates and establishes dissipative coupling. In-phase condensates create an antinode at the center of the waveguide structure that strongly couples to the loss channel. In contrast, when the condensates are in antiphase a node is created at the center and light then only weakly couples to the loss channel. In a symmetric coupler this always leads to the antiferromagnetic coupling depicted in 
.}
{\color{black}In order to achieve a dissipative coupling, a direct exchange of particles between the condensates should be avoided. Instead, we propose to use a potential landscape that realizes a Y-coupling between the condensates. This structure defines a loss channel for the coupled condensate system, i.e. the open end of the Y-coupling, which depends on the relative phase of the two condensates: if the condensates oscillate in-phase, the particle streams emitted from the two condensate interfere constructively in the open channel of the Y-coupling, which maximizes the overall losses of the coupled condensate system. Destructive interference, and thus a minimization of losses, occurs with condensates in antiphase configuration. In this way, the Y-coupling potential leads to a splitting in the imaginary components of the complex energies of the two states. An example of a condensate system coupled in this way is shown in Fig. \ref{fig_cavity}e. The density distribution indicates an overall antisymmetric wavefunction that describes two condensates in antiphase configuration.}
 

\begin{figure}[]
\begin{center}
  \includegraphics[width=\columnwidth]{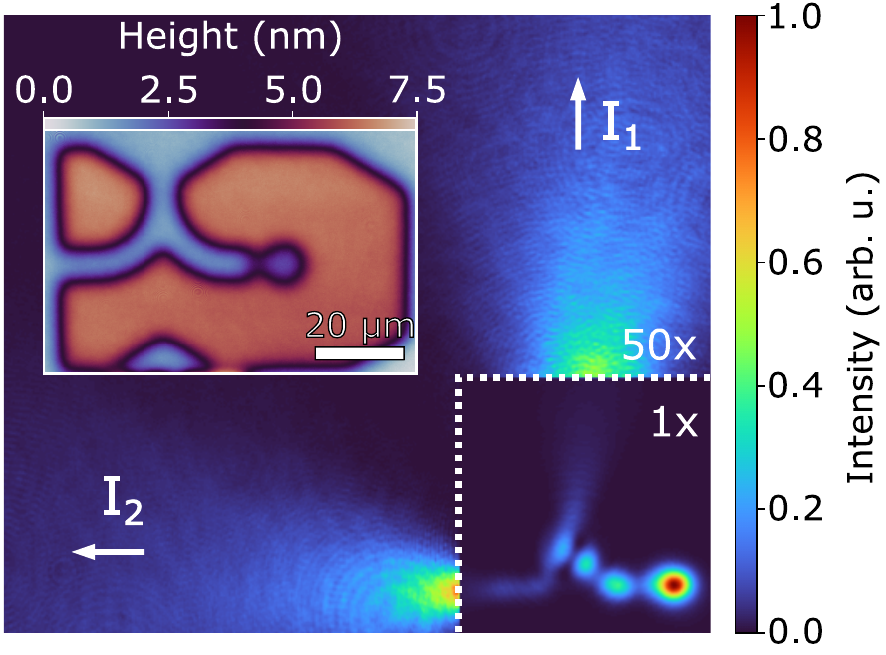}
   \caption{\textbf{$\mid$ Flow of light in the Y-coupler}. A modified Y-coupler structure, as shown in the top left inset figure, is used to determine the splitting between photons flowing out of the structure at the exits (I$_1$, I$_2$). I$_1$ contributes to the dissipative component of the coupling, while I$_2$ contributes a dispersive component. The measured photon density in the area marked with '50x' is multiplied by that factor in order to improve visibility of this low intensity region and highlight the spatial extent of the wavefunction. The measured intensity is averaged over 100 pulses.}

  \label{fig_dis_split}
\end{center}
\end{figure}

It is important to emphasize that\st{neither coupler is strictly dispersive or dissipative in nature.}{\color{black}, strictly speaking, neither the direct coupling method described above is purely dispersive nor the Y-coupling is purely dissipative.} \st{For the direct coupler} {\color{black}Direct particle exchange leads to a splitting of the frequencies of in-phase or antiphase coupled condensates.} \st{t}{\color{black}T}he \st{energy}{\color{black}frequency}-dependent absorption \st{(}governed by the Kennard-Stepanov law\st{)} {\color{black}then} contributes a dissipative component \st{to the coupling} {\color{black}in the case of the direct coupling potential}. \st{For the Y-coupler reflection from the waveguide exit and tunneling through the finite potential barrier at the center of the structure lead to a direct particle exchange between the condensates} {\color{black}For the Y-coupling potential, the particle flow that is emitted by one condensate is not fully directed into the open end of the coupling, but rather partially reaches the other condensate}, which makes the coupling partially dispersive. To quantify this \st{contribution for the Y-coupler} {\color{black}effect,} we {\color{black}can} measure the splitting ratio \st{of the} {\color{black}introduced by the Y-coupling potential, see} \st{Y-coupler geometry in} Fig. \ref{fig_dis_split}. {\color{black}To do that, }\st{T}{\color{black}t}he potential is modified by removing one of the condensate \st{sites} {\color{black}confining potentials} and replacing it by a\st{n} {\color{black}second} open end. This allows for the determination of the splitting ratio by integrating all light leaving the \st{junction} {\color{black}structure} towards either exit and comparing the intensities. \st{We find that $(62 \pm 5) \%$ leaves the structure at the junction $\mathrm{I}_{\mathrm{out}}$.} {\color{black}We find that $(62 \pm 5) \%$ of the particle flow emitted by one condensate leaves the Y-coupling potential through the designated loss channel, while $(38 \pm 5) \%$ of the particles would reach the other condensate.} This light contributes a \st{dissipative} {\color{black}dispersive} component to the coupling. \st{The remaining $38 \pm 5 \%$ enters the waveguide ($\mathrm{I}_\mathrm{wg})$. In the two-condensate system this light would directly couple to the other condensate, hence contributing to the dispersive component of the coupling constant.}

{\color{black}{\bf Critical coupling.}} \st{A key} {\color{black}An important} parameter for spin-glass simulation is the strength of the coupling. \st{First and foremost, t}{\color{black}T}he coupling strength dictates the timescale of the \st{system evolution} {\color{black}synchronization process of the condensate phases}. \st{In other words}{\color{black}In principle}, stronger coupling translates into faster computation times. Stronger coupling can also counteract \st{frequency detuning between condensates (caused by manufacturing defects, for example)} {\color{black}inhomogeneities in the natural (uncoupled) condensate frequencies caused, for example, by manufacturing imperfections leading}
\st{. This leads} to a more robust simulation. {\color{black}The latter is suggested, for example, by the well-known Kuramoto model, which predicts a synchronization process if the coupling strength exceeds a critical value that depends on the natural frequency bandwidth of the oscillators \cite{acebron2005kuramoto}. 

\begin{figure*}[]
\begin{center}
\includegraphics[width=18cm]{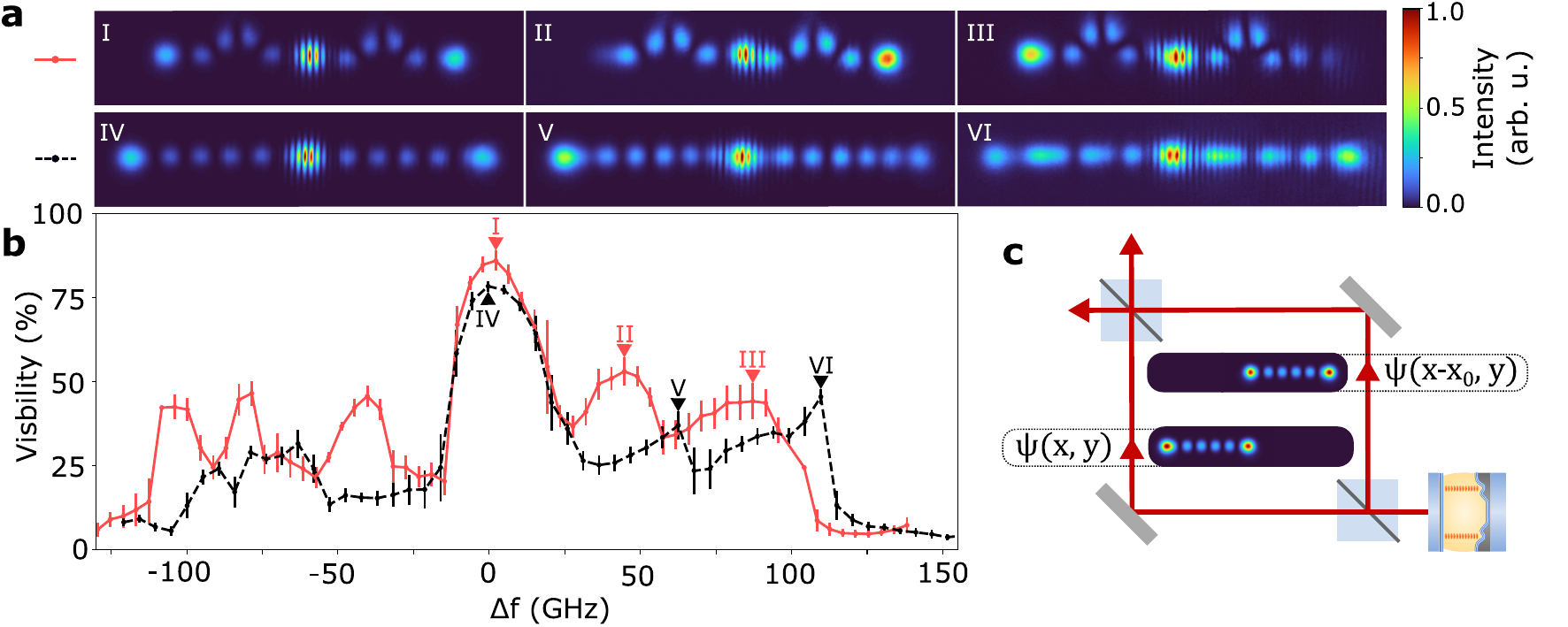}
\caption{\textbf{$\mid$ Determination of the critical coupling.}
To determine the critical coupling, we record the fringe visibility of the interference between the two condensates as function of detuning in the natural frequencies.
\textbf{a} Images of the interference between the coupled-condensates as a function of detuning. To achieve an accurate estimate, we take the mean of the central 5 fringes and average that over 100 pulses.
\textbf{b} Interferometric visibility as a function of detuning between individual condensates.
The Y-coupler (red solid line) and direct coupler (black dashed line) both show several breakdowns and revivals of the visibility. 
Data points I-III mark the maxima of these revivals for the Y-coupler and correspond to the labeled images.
The same holds for data points IV-VI and the direct-coupler curve. 
\textbf{c} Schematic representation of the Mach-Zehnder interferometer used for superposing the condensates and creating the interference pattern.
}
\label{fig_tilt}
\end{center}
\end{figure*}

To demonstrate and quantify critical coupling in systems of coupled photon BECs, we systematically increase the detuning of the natural condensate frequencies and determine the degree of synchronization.} \st{In these coupled oscillator systems the maximum amount of detuning allowed before the system turns incoherent is proportional to the coupling strength. To quantify this, we measure the coherence as a function of the detuning between the condensate sites.} {\color{black}In order to achieve a controlled detuning of the condensate frequencies, we tilt one of the mirrors of the microresonator using piezoelectric actuators.} \st{The light exiting the cavity is first passed through a balanced Mach-Zehnder interferometer before detection on a camera. This interferometer is aligned such that one of the interferometer arms displaces the image, such that the image on the camera contains an overlapping spot of both condensates with interference fringes. The interferometric visibility of this spot is taken as a direct measure for the coherence between the two condensates.} {\color{black}Based on eq. (\ref{teq0}), t}\st{T}ilting the cavity mirrors with respect to each other creates a potential gradient. This gradient results in a potential \st{height} difference between the condensate sites and thus in a detuning between the natural frequencies of the condensates. {\color{black}The experimental quantification of this detuning is by no means trivial. In order to realize this experimentally, we use a potential reconstruction method that we introduced in an earlier work \cite{vretenar2021controllable}. Further details are given in the Supplementary Information.} \st{To calibrate the frequency detuning, we use a potential reconstruction technique, see Supplemental Materials: Tilt Calibration.} {\color{black}The determination of the degree of synchronization, on the other hand, can be realized quite naturally by means of an interferometric measurement. The light transmitted through the cavity mirrors is passed through a balanced Mach-Zehnder interferometer before it is detected on a camera. The interferometer is intentionally misaligned such that the two condensate wavefunctions overlap and interfere on the camera creating a stripe pattern. The interferometric visibility of this pattern is taken as a direct measure for the coherence between the two condensates.}

\st{In Fig. 
coherence remains high for both coupler designs with a detuning $<10\,$GHz. Rather than a uniform drop in coherence there are recurrences in the coherence after this point. These coincide with states close to resonance, as shown by the snapshots of the wavefunctions in the figure. The direct coupler shows sharper features in the visibility curve compared to the Y-coupler . This is expected, since the additional losses in the Y-coupler can cause a broadening of resonance peaks and a smoothing of the overall curve. The intensity patterns from the experiment hint towards a sign change in the first-order coherence of the condensates. This is confirmed by simulations with a dissipative Schr\"odinger equation that describes the dynamics of the photon Bose-Einstein condensate system, see Supplemental Materials: Theoretical Methods.}
{\color{black}In Fig. 3 we show experimental results that describe the coherence of two coupled Bose-Einstein condensates as a function of detuning of their natural frequencies. The results suggest that the phase coupling of the condensates can withstand a detuning of up to 100 GHz until it is finally lost and the condensates start to oscillate independently. It turns out, however, that the coherence for both the direct coupling potential and the Y-coupling potential is lost in a non-monotonic way, i.e. several breakdowns and revivals in the visibility of the condensate interference can be observed. The experimentally obtained mode patterns suggest that this non-monotonic behaviour could be related to changes between in-phase and antiphase correlations with increasing detuning, i.e. sign changes in the first-order correlations $g^{(1)}$ of the two condensates. Indeed, a linear potential gradient is expected to change the phase delay that the photons collect as they propagate in the region between the condensates, which could be the cause for such sign changes. However, this interpretation cannot be conclusively confirmed on the basis of our experimental data alone. This is mainly due to the fact that our experiments, as they are currently being carried out, only determine the visibility of the interference, i.e. $\vert g^{(1)} \vert$, but do not reveal the sign in the first-order correlations. To investigate this situation further, we carried out numerical simulations that allow us to access the underlying $g^{(1)}$-function. Further details are given in the Supplementary Information. The results of these simulations indeed support the above interpretation that the breakdowns and revivals of the visibility are caused by sign changes in $g^{(1)}$. 

The visibility in Fig. \ref{fig_tilt} is shown as function of both positive and negative detunings. In principle, one would expect that the graphs should be fully symmetric with respect to zero detuning - which is, however, not the case. This results from the axis of rotation being offset compared to the center of the structure. In this way, a common frequency offset is introduced for both condensates, which is not reflected in the value of the detuning, but which can be different for positive and negative detunings and thus breaks the symmetry of the measurement curve to some extent.}
\st{Another key difference is the higher initial coherence of the Y-coupler compared to the direct coupler. This is also present in the simulations. The effect is most likely due to the higher dissipation of the symmetric state, suppressing its contribution to final time-averaged state.}

\begin{figure}[]
\begin{center}
  \includegraphics[width=\columnwidth]{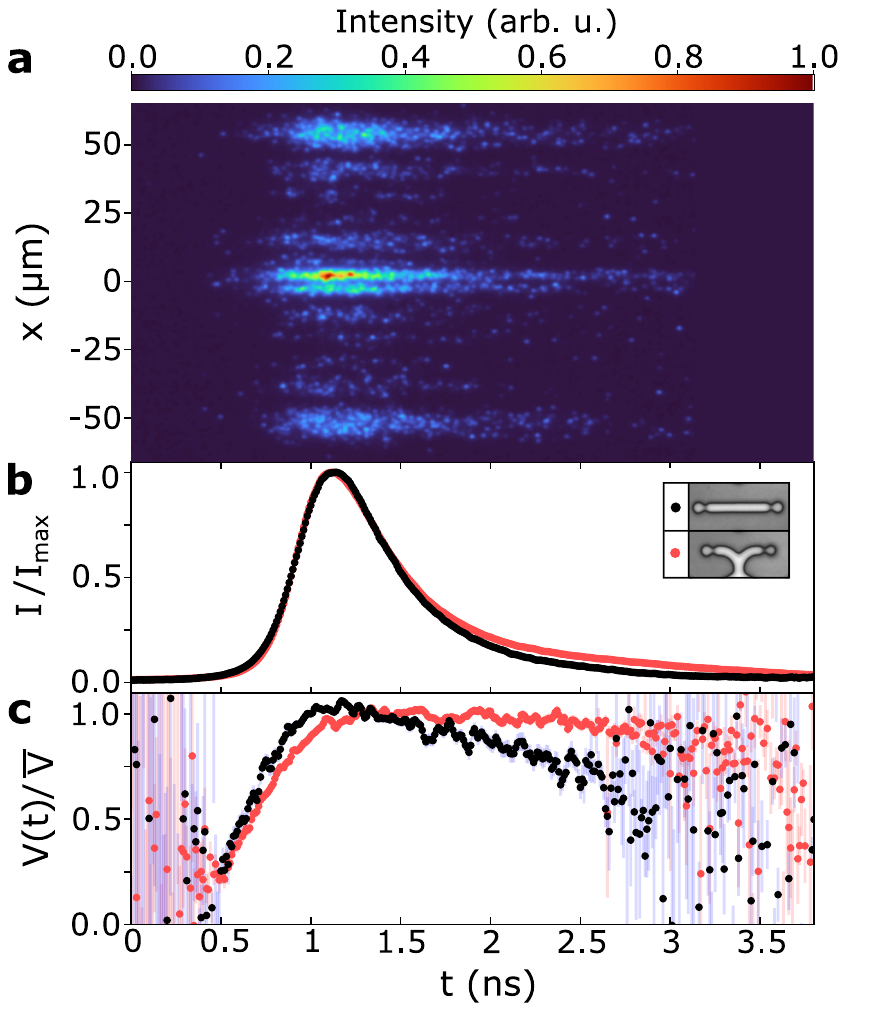}
   \caption{\textbf{$\mid$ Time-resolved coherence and condensate population for the direct and Y-coupler}. \textbf{a} Streak camera image of a condensation in the Y-coupler. This is a time-resolved measurement of a cross section along the coupler. The central spot contains interference between both condensates, the visibility of these fringes over time is measured as $V(t)$. Due to (shot) noise in the measurement, averaging over many pulses is required. We postselect the pulses to remove formation jitter and intensity fluctuations. This leaves 247 and 757 out of 4000 images for the direct coupler and Y-coupler, respectively.  \textbf{b} Averaged intensity envelopes of the direct coupler (black) and the Y-coupler (red), normalized to the maximum intensity. Both curves are very similar in envelope, the slightly longer tail in the pulse of the Y-coupler most likely stems from pump pulse fluctuations present in a subset of all streak camera images. \textbf{c} Fringe visibility of the interference between the two condensates for the direct coupler (black) and Y-coupler (red), visibility is normalized to the time-averaged visibility of the integrated pulse. Error bars show the standard error for all data points. }

  \label{fig_coh_time}
\end{center}
\end{figure}

{\color{black}{\bf Time-resolved synchronization process.}} In a second set of experiments, we \st{want to} investigate the condensation and synchronization process in a time-resolved manner using a streak camera. 
{\color{black}As before, we let the light from the cavity pass through an interferometer in order to make the coherence of the two coupled condensates visible as an interference pattern. This interference pattern is imaged onto the entrance slit of a streak camera so that we can examine it in a time-resolved manner after exciting the cavity with a nanosecond optical pulse. An example of such a measurement is shown in Fig. \ref{fig_coh_time}a. Here, the light intensity is shown along a cross section through the interference pattern as a function of time. In the case shown, two interference fringes are visible (located within a few micrometer around $x=0$), which can be analyzed to determine the fringe visibility.}
\st{Fig.}
{\color{black}Figures \ref{fig_coh_time}b,c} show\st{s} the \st{average normalized} intensity and \st{normalized} fringe visibility \st{of the pulses} as a function of time {\color{black}(when averaging over several hundred excitation pulses)}. \st{Since the rise time is dominated by the gain in the medium and fall time is predominantly determined by the lifetime of the photons in the cavity, the intensity profile of both pulses is similar. The fringe visibility envelope shows two main differences between the designs.} {\color{black}From these measurements it can be concluded that the time evolution of the condensate population is rather similar for the two types of coupling (red and black curves).} \st{First, the rise time of the visibility (based on a linear fit in between $20-95\%$ of the rising edge) is different for each coupler design. The rise times of the Y- and direct coupler coherence is $(570 \pm 10)\,$ps and $(440 \pm 10)\,$ps, respectively. The other key difference is the fringe visibility drop off. The visibility for the direct coupler clearly decreases in the tail of the pulse while the Y-coupler remains approximately constant.} {\color{black}Differences arise, however, in the time evolution of the visibility. Here, our measurements show that the coherence is established somewhat faster with the direct coupling (rise time $(440 \pm 10)\,$ps) than with the Y-coupling (rise time $(570 \pm 10)\,$ps). On the other hand, almost full coherence is retained with the Y-coupling even when the condensate population drops again, while the coherence in the direct coupling potential decreases more quickly.}

\st{The next set of experiments assess the effects of asymmetric pumping on the condensation process. Here, high pump intensities were used to increase signal level and only one condensate site is pumped by the excitation laser. The photon density in the other condensate originates from direct particle exchange. The pump spot is aligned such that the systems condense into a  stable coherent state. We measure the population imbalance between the two condensates over the observed part of the pulse as well as the delay between the condensate formation between both sites. The delay is measured as follows: First the intensity of both condensates is normalized to their respective maximum values. Then a trigger point is chosen at half of this maximum value and the timing difference between both trigger points is defined as the delay. 
contains an example of a typical pulse for each coupler design as well as a normalized intensity curve for both condensate sites as function of time. The measurements are ensemble averaged over multiple pulses. The population imbalance is measured by integrating the intensity of both condensates over the duration of the pulse. The population imbalance for the direct coupler is 1.17 $\pm$ 0.06 and the delay is 9 $\pm\,$2 ps. The Y-coupler has a population imbalance of 2.78 $\pm$ 0.09 and a delay of 36 $\pm\,$2 ps. These measurements clearly indicate that the Y-coupler design has a significant dispersive component, given by the fact that at least some direct particle exchange is required to populate both condensates under single-site illumination. However, both the population imbalance and the formation delay are much larger for the Y-coupler case, leading to the conclusion that the particle exchange, and therefore the dispersive coupling, is significantly weaker for this design.}

{\color{black}In general, spin-glass simulation based on networks of photon or polariton Bose-Einstein condensates not only requires precise control of the physical couplings between the condensates, it is also necessary to control the condensate populations, since these represent scaling factors for the effective coupling in the simulated XY model \cite{vretenar2021controllable,kalinin2018global}. To create a homogeneous network of condensates that all have the same population, it can generally be necessary to distribute the optical gain inhomogeneously over the network. A natural question that arises is how the two types of coupling examined in this work behave with regard to inhomogeneous or unbalanced optical pumping. In contrast to the experiments carried out so far, we will concentrate the optical gain to a single condensate location in the following measurements to answer this question.

\begin{figure}[]
\begin{center}
  \includegraphics[width=\columnwidth]{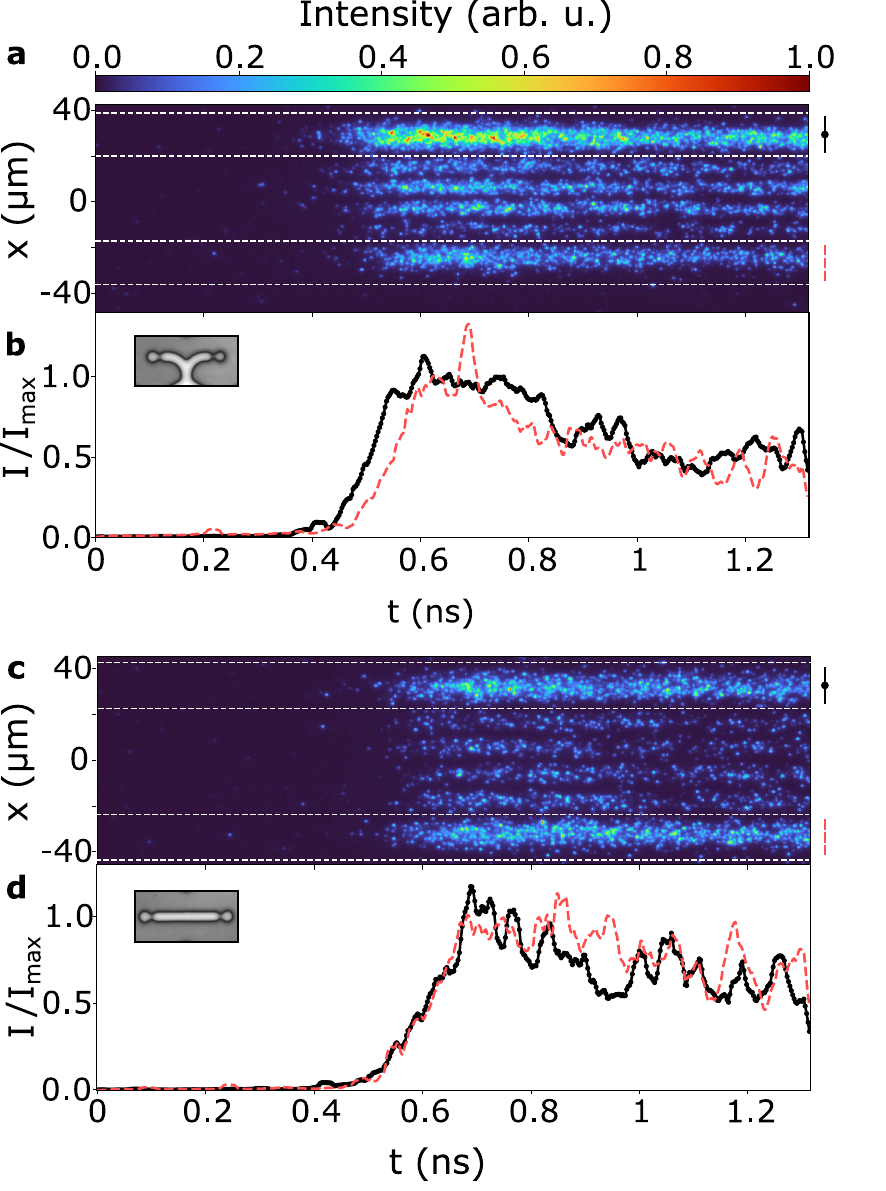}
    \caption{\textbf{$\mid$ Time delay in condensation under imbalanced optical pumping.}  \textbf{a} Streak camera image of the Y-coupler condensate under single-site illumination. \textbf{b} The spatially integrated intensity of the streak camera image regions marked by the white-dashed lines, where the condensates are located. The black line corresponds to the optically pumped condensate. A clear delay in intensity build up is apparent in both figures. Both intensities are normalized to the average of the 50 maximum values. \textbf{c,d} The same measurement, as in \textbf{a,b}, for a condensation in the direct coupler structure. For the determination of the delay and populations imbalance, the values are ensemble averaged over 38 pulses for the direct coupler and 99 pulses for the Y-coupler. The reason for the difference in pulse numbers is the necessity of postselection due to the formation jitter.}
  \label{fig_delay}
\end{center}
\end{figure}

Figure 5 shows two examples for the time evolution of the condensate populations in the case of direct coupling (Fig. 5a,b) and Y-coupling (Fig. 5c,d). Although only one condensate location at $x=30\upmu$m experiences optical gain in these experiments, a second condensate at $x=-30\upmu$m forms for both coupling types, the population of which is fed solely through particle exchange with the other condensate. In the case of direct coupling (Fig. 5c), this even goes so far that both condensate populations are almost identical (population ratio 1.17). Although this behavior may seem counterintuitive at first, an equal population of the condensates is indeed theoretically expected if the coupling rate between the condensates is sufficiently high \cite{vretenar2021controllable}. In case of the Y-coupling potential (Fig. 5a), the direct particle exchange is deliberately suppressed, so that, as expected, significantly different condensate populations result (population ratio 2.78). Further differences between the two types of coupling arise in the temporal dynamics of the condensation process. To investigate this, we determine the normalized condensate population $I/I_\text{max}$ for both condensates, see Fig. 5b,d. While both condensates are formed almost synchronously for the case of direct coupling (time delay of $(9 \pm\,$2)\,ps), the formation of the second condensate has a considerable time delay of $(36 \pm\,$2)\,ps in case of the Y-coupling potential.} 

\st{It is also possible to get a coherent superposition of the symmetric and antisymmetric state of the direct coupler. This is shown in 
. Here, the condensate is pumped slightly off-center from the trapping potential. The measurement clearly shows a beat frequency of $(14.5 \pm 0.5)\,$GHz between the symmetric mode and the antisymmetric. This gives a direct measure for the dispersive component of the coupling in the direct coupler. It is also clear that oscillation amplitude decreases over time and the system converges to an antisymmetric mode with 4 fringes. The temporal evolution of the coupled condensates is an interplay of many different mechanisms. Thermalization can drive the system to a lower energy state as is the case for this particular measurement, similar to previous measurements on excited states of single photon BECs
. However, effects caused by gain redistribution through repeated absorption and emission cycles can also influence the process. The coupling strength found here is consistent with the first drop in coherence in 
. In fact, between points I and II of 
, the direct coupler displays integrated intensity patterns similar to those present in 
and it is therefore in the same oscillatory state. The potential gradient, present in that case, leads to a mapping of the phenomenon onto the DC-Josephson effect, as was done previously in polaritonic systems 
. For the Y-coupler, this frequency measurement is not possible due to the much higher loss for the symmetric state in that case.}

\begin{figure}[]
\begin{center}
  \includegraphics[width=\columnwidth]{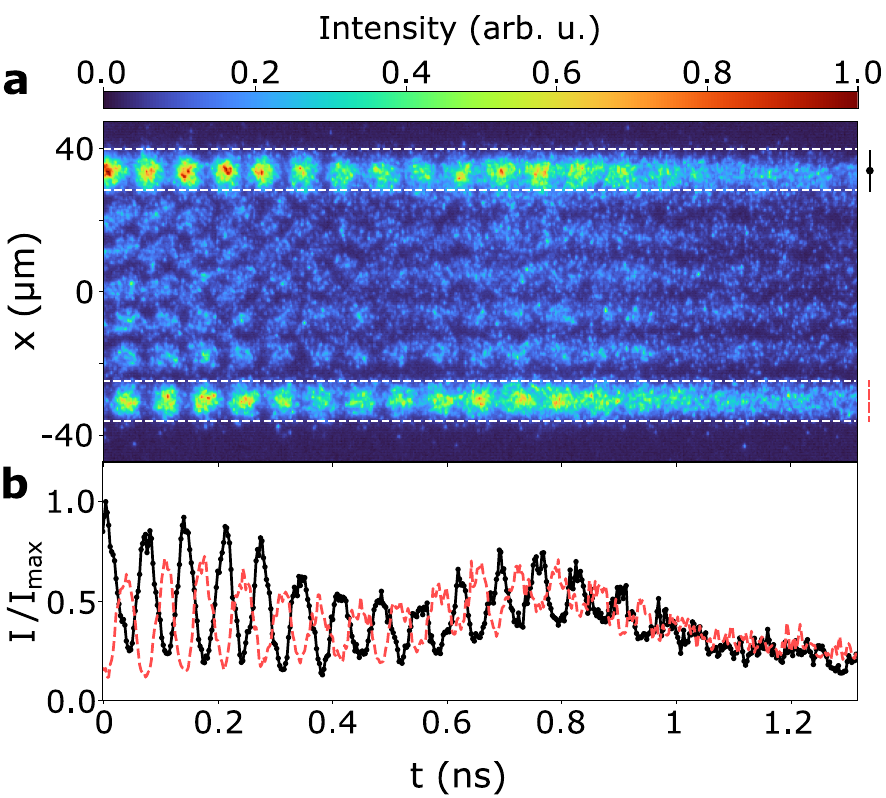}
   \caption{\textbf{$\mid$ Coherent oscillation under imbalanced optical pumping of the direct coupler.} \textbf{a} Streak camera image of the direct coupler with off-center pumping of a single condensate (compared to the confining potential). \textbf{b} Integrated intensity of the individual condensate sites. An oscillation is present with a frequency of $(14.5 \pm 0.5)\,$GHz resulting from the coherent superposition of the antisymmetric state (with 6 nodes) and the symmetric state (with 5 nodes). The oscillation is damped as the system evolves from a superposition of both states into the single lower-energy antisymmetric state.}

  \label{fig_osc}
\end{center}
\end{figure}

{\color{black}A substantially different dynamical behavior than shown in Fig. 5 is observed when the system is set in a parameter range between two stable configurations. The latter can be achieved, for example, by introducing a small mismatch between the location of the optical pumping and the confining potential of the condensate, as is done in the measurement shown in Fig. 6. In this case, we observe that the system tends to go through oscillations in the condensate populations. These oscillations are caused by superposition states that arise when two (or more) different states experience similar optical gain and thus both are realized simultaneously by the system \cite{ding2019dispersive}. In the example of Fig. 6, this is a superposition of a symmetric wavefunction with 6 nodes and an antisymmetric wavefunction with 5 nodes. The observed oscillation frequency of $14.5 \pm 0.5\,$GHz can be considered as the difference between the real parts of the complex energies of these two states. Interestingly, it turns out that this superposition state is not stable. For longer periods of time, the antisymmetric state starts to dominate, so that the oscillations in the condensate populations eventually disappear. It seems plausible to consider this as part of a thermalization process in which the state with the lower energy prevails due to the lower reabsorption by the optical medium. A similar phenomenon was previously observed in harmonically trapped photon gases \cite{schmitt2015thermalization}. We would like to emphasize that the oscillative behavior shown in Fig. 6 is quite typical for the direct coupling of two condensates. In particular, it also occurs in the case of symmetrical optical pumping. Both the experimentally determined photon densities and our numerical simulations suggest that superposition states and oscillative behavior occur at the breakdowns of visibility in the tilting experiment in Fig. 3. \st{Since potential gradients are present in this case, the oscillative behavior here can be interpreted as the DC Josephson effect}
}

\section*{Discussion}\label{appExpM}
In this work we investigated the dynamics of a two-condensate system with dissipative and dispersive couplings. One of the most interesting applications of such a system is to use it as a building block for an \st{all-optical} {\color{black}analog} spin-glass simulator. \st{We find that proper excitation of the condensate sites is crucial for fast settling into the low energy state. For balanced excitation, such as shown in Figure 
, the condensation process quickly realizes the final state. On the other hand, under deliberately imbalanced excitation like Figure 
, the evolution towards the intended low energy state can take $> 1\,$ns for the direct coupler.}
{\color{black}Various approaches to realizing such simulators are currently being investigated. In the different approaches there are also different challenges that have to be overcome, for example, the replacement of digital electronic components in the computation process, miniaturization, non-local couplings, and scaling. For the photon BEC system, scaling to larger system sizes is the most challenging aspect. The larger the network of photon Bose-Einstein condensates, the larger is the bandwidth of the natural (uncoupled) condensate frequencies, since even the best resonator mirrors are not perfectly flat. In principle, this can lead to phase coherence only being established for subsystems, but not for the system as a whole. The good news, however, is that the coupling of the condensates is relatively strong and thus also quite robust against detuning caused by mirror imperfections. From Fig. 3 it can be seen, for example, that a detuning of 10 GHz in the natural frequencies only leads to a minor decrease in coherence in the coupled system. Converted into a height difference, a detuning of 10 GHz corresponds to a height difference of 0.2 nm (assuming a resonator length of 10 $\upmu$m). This means that the mirrors used should not have any height deviations greater than 0.2 nm in an area of, for example, several square millimeters in order to sufficiently restrict the bandwidth of natural frequencies in the BEC network. This is at the limit of what is technically feasible today, but it seems achievable. The direct-laser-writing technique we use for nanostructuring mirror surfaces does indeed achieve this level of accuracy. It is furthermore possible to actively control the condensate frequencies using a thermo-responsive optical medium \cite{dung2017variable}.}

\st{As mentioned above, one of the key parameters for determining the viability of a spin-glass simulator is the coupling strength. In our system, high coherence is maintained for detuning smaller than 10\,GHz. Beyond that, a lower level of coherence is maintained for $> 100\,$GHz. This indicates that the coupling strength is at least in the tens of GHz range.  This also limits the maximum amount of site-to-site potential height variation to $\Delta d \approx 0.2\,$nm for a 10\,$\upmu$m cavity length. This is feasible with our current manufacturing technique. Furthermore, actively tuning the potential, for example through a thermosensitive polymer solution 
, can also easily achieve these levels of potential height variation. It should also be noted that in this work no special care was taken to optimize this coupling which definitely leaves room for optimization.}

\st{Even though the coupling strengths are similar, there are also significant differences between both coupler designs. The Y-coupler suppresses even modes and therefore does not show the strong long term oscillations that can be present in the direct coupler under unfavourable conditions. Another clear difference is the fact that the direct coupler shows only a minor population imbalance under changes in illumination while the Y-coupler shows a much larger imbalance. This is attributed to a much higher particle exchange in the direct coupler. If direct control over the individual population is required, then the direct coupler with strong coupling is not suitable for this application. This was indicated by previous theoretical results which show that, for sufficiently strong dispersive coupling, populations in both condensates become equal to maximize gain 
. The main drawback of the Y-coupler is the increased design complexity compared to the direct coupler, since for more complex structures high loss areas need to be created to dissipate the light.}

{\color{black}Mapping a spin model Hamiltonian to the gain function of an optical spin-glass simulator is a prerequisite for the simulator to start sampling low-energy configurations. It is furthermore desirable that the dynamics of the simulator have fixed points (including the ground state of the simulated spin model). In the photon BEC system and related platforms this is only the case if the dispersive part of the physical couplings is sufficiently small \cite{kalinin2018global}. Dissipative types of coupling, as proposed and investigated in this work, are particularly desirable in this regard. Another requirement is the ability to control the condensate populations individually. Here, interesting differences arise between the two types of coupling investigated in this work. As shown in Fig. 5, direct coupling tends to balance the populations of coupled condensates even with unbalanced optical gain. This property can be useful in specific cases, for example, for simulations of spin models on $N$-regular graphs \cite{vretenar2021controllable}. In general, however, one would like the condensate population to be precisely controllable by the provided optical gain, which is indeed possible with the Y-coupling, since it keeps direct particle exchange sufficiently low. For the future, it seems both possible and desirable to combine dispersive and dissipative types of coupling by controlling the flow of photons between the condensates, for example, with the help of refractive index changes \cite{vretenar2021controllable}. This would extend the class of physical models that can be investigated using analog simulations.}

\section*{Acknowledgements}
\vspace{3mm}
This work was supported by the NWO (grant no. OCENW.KLEIN.453).

\section*{Methods}\label{appExpM}

{\bf Microcavity system.} The microcavity is formed from two high-finesse mirrors, one of which is nano-structured using a direct-laser-writing method, see Ref. \cite{kurtscheid2020realizing}. The height profiles of the mirror surfaces are determined via Mirau interferometry. We use a commercially available interferometric microscope objective for this purpose (20X Nikon CF IC Epi Plan DI). The large free spectral range of the cavity, combined with the limited gain bandwidth of the dye, causes the longitudinal mode number to be a conserved quantity under repeated absorption and emission cycles. We work at mirror separations between $D \simeq 5 \, \mathrm{\upmu m}$ and $D \simeq 10 \, \mathrm{\upmu m}$ for the time-averaged measurements and $D \simeq 30 \, \mathrm{\upmu m}$ for time-resolved measurements. The benefits of working at these comparatively large mirror separations are increased signal strength and reduced sensitivity to surface imperfections.

$\,$

{\bf Optical medium.} The dye in the microcavity consists of 10\,mmol/L rhodamine 6G dissolved in ethylene glycol. We use an optical parametric oscillator with a 5\,ns pulse duration at 490\,nm to optically excite the dye. 

$\,$

{\bf Time-resolved measurements.} For the time-resolved measurements, we use the streak camera Hamamatsu C10910-01 with a M10913-01 slow single sweep unit. 

$\,$

Further experimental methods are presented in the Supplementary Information.

\bibliography{references.bib}
\end{document}